\documentclass[aps,twocolumn,preprintnumbers,amsmath,amssymb,nofootinbib,superscriptaddress,notitlepage]{revtex4}
\usepackage{graphicx,color,dcolumn,booktabs,bm}
\usepackage{longtable,lscape}
\usepackage{txfonts}
\usepackage{overpic}
\usepackage{amssymb}
\usepackage{indentfirst}
\usepackage{feynmf}
\usepackage{slashed}
\usepackage{cases}
\usepackage{multirow}
\usepackage{appendix}
\usepackage[figuresright]{rotating}
\usepackage[colorlinks,
citecolor=blue,
anchorcolor=red,
menucolor=red,
linkcolor=red,
filecolor=red,
runcolor=red,
urlcolor=blue,
frenchlinks=red]{hyperref}
\usepackage{epstopdf}
\usepackage{bm}
\usepackage{tabularx}
\usepackage{threeparttable}
\usepackage{bbding}
\usepackage{array}
\usepackage[section]{placeins}
\usepackage{makecell}
\usepackage{comment}
\makeatletter

\newcommand{\Rmnum}[1]{\expandafter\@slowromancap\romannumeral #1@}
\makeatother

\usepackage{soul}
\begin{document}
	\title{Reevaluating the $\psi(4160)$ Resonance Parameter Using $B^+\to K^+\mu^+\mu^-$ Data in the Context of Unquenched Charmonium Spectroscopy}	
	\author{Tian-Cai Peng}
	\email{pengtc20@lzu.edu.cn}
	\author{Zi-Yue Bai}
	\email{baizy15@lzu.edu.cn}
	\affiliation{School of Physical Science and Technology, Lanzhou University, Lanzhou 730000, China}
	\affiliation{Lanzhou Center for Theoretical Physics, Key Laboratory of Theoretical Physics of Gansu Province, Key Laboratory of Quantum Theory and Applications of MoE and Gansu Provincial Research Center for Basic Disciplines of Quantum Physics, Lanzhou University, Lanzhou 730000, China}
	\affiliation{Research Center for Hadron and CSR Physics, Lanzhou University and Institute of Modern Physics of CAS, Lanzhou 730000, China}
	
	\author{Jun-Zhang~Wang}
	\email{wangjzh2022@pku.edu.cn}
	\affiliation{School of Physics and Center of High Energy Physics, Peking University, Beijing 100871, China}
	
	\author{Xiang~Liu\footnote{Corresponding author}}
	\email{xiangliu@lzu.edu.cn}
	\affiliation{School of Physical Science and Technology, Lanzhou University, Lanzhou 730000, China}
	\affiliation{Lanzhou Center for Theoretical Physics, Key Laboratory of Theoretical Physics of Gansu Province, Key Laboratory of Quantum Theory and Applications of MoE and Gansu Provincial Research Center for Basic Disciplines of Quantum Physics, Lanzhou University, Lanzhou 730000, China}
	\affiliation{Research Center for Hadron and CSR Physics, Lanzhou University and Institute of Modern Physics of CAS, Lanzhou 730000, China}
	\affiliation{MoE Frontiers Science Center for Rare Isotopes, Lanzhou University, Lanzhou 730000, China}

	\date{\today}
	
	\begin{abstract}

	A puzzling phenomenon, where the measured mass of the $\psi(4160)$ is pushed higher, presents a challenge to current theoretical models of hadron spectroscopy. This study suggests that the issue arises from analyses based on the outdated quenched charmonium spectrum. In the past two decades, the discovery of new hadronic states has emphasized the importance of the unquenched effect. Under the unquenched picture, six vector charmonium states—$\psi(4040)$, $\psi(4160)$, $\psi(4220)$, $\psi(4380)$, $\psi(4415)$, and $\psi(4500)$—are identified in the $4 \sim 4.5$ GeV range, contrasting with the three states predicted in the quenched model. We reevaluate the resonance parameters of the $\psi(4160)$ using the di-muon invariant mass spectrum of $B^+ \to K^+ \mu^+ \mu^-$ and unquenched charmonium spectroscopy. Our analysis indicates previous experimental overestimations for the mass of the $\psi(4160)$. This conclusion is supported by analyzing $e^+e^- \to D_s \bar{D}_s^*$, which finds the $\psi(4160)$ mass at $4145.76 \pm 4.48$ MeV. Our findings have significant implications for both hadron spectroscopy and search for new physics signals by $R_K$.
		
	\end{abstract}
	
	\maketitle

\section{Introduction}
Since the discovery of the first charmonium state ($J/\psi$ particle) \cite{E598:1974sol,SLAC-SP-017:1974ind}, fifty years have passed. Over this time, dozens of charmonium states have been documented in the Review of Particle Physics (RPP) \cite{ParticleDataGroup:2024cfk}. Examining the experimental data for these observed charmonium states reveals an intriguing phenomenon associated with the charmonium $\psi(4160)$. Notably, its reported mass as a resonance parameter has shifted over the years, increasing from the initially measured $\sim4160$ MeV to the more recent value of $\sim4190$ MeV.

\begin{figure}[htbp]\centering
	    \includegraphics[width=0.45\textwidth]{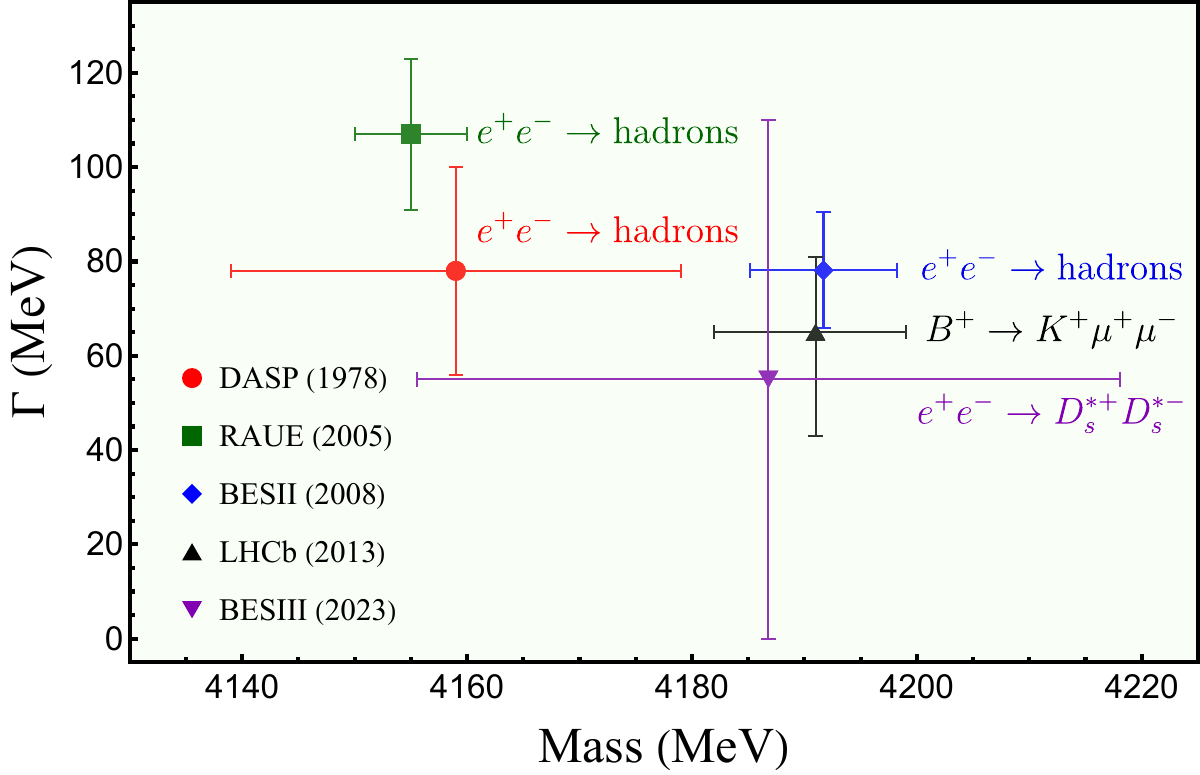}
		\caption{Resonance parameters of the $\psi(4160)$ measured by different experiments~\cite{DASP:1978dns,Seth:2004py,BES:2007zwq,LHCb:2013ywr,BESIII:2023wsc}, expressed in MeV.}
		\label{fig0}
	\end{figure}

The $\psi(4160)$ was first discovered in 1978 by the DASP Collaboration, which measured its mass to be $4159 \pm 20$ MeV \cite{DASP:1978dns}. This finding was later confirmed by the Crystal Ball Collaboration, which reported a mass of $4155 \pm 5$ MeV \cite{Seth:2004py}. Prior to 2008, experimental data consistently placed the mass of the $\psi(4160)$ at approximately 4160 MeV, as shown in Fig. \ref{fig0} and supported by earlier studies \cite{DASP:1978dns,Seth:2004py}. This estimate led to the state being designated as $\psi(4160)$. However, in 2008, an analysis of the total hadronic cross-section, $\sigma(e^+e^-\to \text{hadrons})$, conducted by the BESII Collaboration, challenged this mass determination \cite{BES:2007zwq}.
In 2013, the LHCb Collaboration provided further results by analyzing the decay $B^+\to K^+ \mu^+\mu^-$. Their measurements indicated a mass of $4191^{+9}_{-8}$ MeV for the $\psi(4160)$ \cite{LHCb:2013ywr}, corroborating the findings of the BESII analysis. Consequently, the Particle Data Group (PDG) updated the average mass of the $\psi(4160)$ to $4191 \pm 5$ MeV \cite{ParticleDataGroup:2024cfk}. In fact, this shift in the mass of the $\psi(4160)$ poses a challenge to current studies of charmonium spectroscopy, as the $\psi(4160)$ is often used as a reference point in phenomenological models.

We should revisit several experimental analyses that have contributed to this observed mass shift. A common approach has been to apply the mass spectrum of charmonium derived from quenched potential models to these analyses. In the energy range of $4 \sim 4.5$ GeV, three vector states—$\psi(4040)$, $\psi(4160)$, and $\psi(4415)$—are typically identified.
However, it is crucial to emphasize that we are now in an era of high-precision hadron spectroscopy, where unquenched effects cannot be overlooked and must be adequately considered in the interpretation of hadron spectra. Studies from the past decades \cite{He:2014xna,Chen:2014sra,Chen:2015bma,Chen:2017uof,Wang:2019mhs,Wang:2022jxj,Wang:2023zxj,Peng:2024xui} suggest the existence of six vector charmonium states in the mass range of $4 \sim 4.5$ GeV: $\psi(4040)$, $\psi(4160)$, $\psi(4220)$, $\psi(4380)$, $\psi(4415)$, and $\psi(4500)$. This scenario markedly differs from the predictions of the quenched potential model.
Given this perspective, it is necessary to reevaluate the resonance parameters of the $\psi(4160)$ by incorporating the refined mass spectrum of vector charmonium from the unquenched framework alongside experimental data from the decay $B^+ \to K^+ \mu^+ \mu^-$. In this process, the $\mu^+ \mu^-$ pair can be attributed to contributions from vector charmonium states.

\section{Analysis of di-muon mass spectrum}	
Since the initial experimental observation of the $\psi(4160)$, its reported mass has varied over time. After the mass shifted to a higher value around 4190 MeV, the $\psi(4220)$ state was subsequently observed in various processes~\cite{BaBar:2005hhc,BESIII:2016bnd,BESIII:2020oph,BaBar:2006ait,BESIII:2017tqk,BESIII:2016adj,BESIII:2019gjc,BESIII:2022joj,BESIII:2018iea,BESIII:2023tll}. In our previous work~\cite{Wang:2019mhs}, we described both of these states using a unified unquenched charmonium spectrum. This spectrum has been tested in several strong decay processes~\cite{Wang:2019mhs,Wang:2022jxj,Wang:2023zxj,Peng:2024xui}, and the interference effect between the $\psi(4160)$ and $\psi(4220)$ has proven to be crucial. Therefore, when discussing the resonance parameters of the $\psi(4160)$, it is essential to account for the presence of the $\psi(4220)$.

	
	To investigate this issue, in this work, we first analyzed the di-muon invariant mass spectrum in the $B^+ \to K^+ \mu^+\mu^-$ decay under the unquenched charmonium spectrum. According to the experimental analysis, the mass of $\psi(4160)$is about 4190 MeV. For describing the contribution of charmonium in the $\mu^+\mu^-$ mass spectrum in the $B^+ \to K^+ \mu^+\mu^-$ process, the probability density function of the signal should be proportional to the square of the total amplitude, i.e.,

	\begin{align}		
		\mathcal{P}_{\mathrm{sig}}\propto |\mathcal{A}_{\mathrm{Tot}}|^{2}.
	\end{align}	
 
 {The decay process $B^+\to K^+\mu^+\mu^-$ can occur through three distinct mechanisms, where the dimuon pair ($\mu^+\mu^-$) couples to a $Z^0$ boson, a photon ($\gamma$), or a vector resonance. These contributions are represented by the amplitudes $\mathcal{A}_{\mathrm{non-res}}^{\mathrm{AV}}$($Z^0$), $\mathcal{A}_{\mathrm{non-res}}^{\mathrm{V}}$($Z^0$ and $\gamma$), and $\mathcal{A}_{\mathrm{res}}^{n}$. The subscripts $\mathrm{AV}$ and $\mathrm{V}$ are used to denote the first two terms, reflecting the axial-vector ($\mathrm{AV}$) and vector ($\mathrm{V}$) nature of the couplings involved.} The total amplitude is
	\begin{align}		
		\mathcal{A}_{\mathrm{Tot}}=\mathcal{A}_{\mathrm{non-res}}^{\mathrm{AV}}+\mathcal{A}_{\mathrm{non-res}}^{\mathrm{V}}+\sum_{n}f_{n}e^{i\delta_{n}}\mathcal{A}_{\mathrm{res}}^{n},
	\end{align}	
	
	In this work, we focus solely on the vector charmonium states as the resonant contributions to the dimuon mass distribution. Consequently, only the vector component of the non-resonant amplitude will interfere with the resonance amplitude. { At the energy region around 4 GeV, the axial vector contributions of $Z^0$ boson to the signal is a constant \cite{LHCb:2013ywr}. Therefore, the contribution of the axial vector coupling will not affect our description of the line shape, so we ignore it in our analysis.} The probability density function of the signal can therefore be expressed as,
	
	\begin{align}	
		\mathcal{P}_{\mathrm{sig}}\propto\left|\mathcal{A}_{\mathrm{non-res}}^{\mathrm{V}}+\sum_{n}f_{n}e^{i\delta_{n}}\mathcal{A}_{\mathrm{res}}^{n}\right|^{2}.
	\end{align}	
	The non-resonance contribution can be parameterized by an Argus function~\cite{ARGUS:1990hfq}, expressed as $\mathcal{A}_{\mathrm{non-res}} = m_{\mu^+\mu^-} u^{p/2} e^{q m_0 / 2}$, where $u = 1 - m_{\mu^+ \mu^-} / m_0^2$, and $m_{\mu^+ \mu^-}$ is the invariant mass of the di-muon. The parameters $p=1.579$, $q=-2.861$ MeV$^{-1}$ and $m_0=4780$ MeV that can be determined by fitting to experimental data.
	
	For the resonance contribution, we use a phase space factor corrected by the Breit-Wigner distribution, i.e.,

	\begin{align}	
		\mathcal{A}_{\mathrm{res}}=\mathrm{phsp}(m_{\mu^+\mu^-})\frac{m_R\Gamma_R}{(m_R^2-m_{\mu^+\mu^-}^2)-im_R\Gamma_R},
	\end{align}	
	where $\mathrm{phsp}(m_{\mu^{+}\mu^{-}}) = \frac{1}{2m_{B^{+}}} \lambda(m_{B^{+}}^{2},m_{K^{+}}^{2},m_{\mu^{+}\mu^{-}}^{2})^{1/2}$ with $\lambda(a,b,c) = a^2 + b^2 + c^2 - 2(ab + ac + bc) $ is the phase space of $B^+ \to K^+ + R$.

		\begin{table}[t]
		\renewcommand\arraystretch{1.5}
		\caption{The masses and widths of higher charmonium states in the range of $4.0\rm{-}4.5$ $\text{GeV}$, which were obtained from the theoretical predictions \cite{Wang:2019mhs, Wang:2022jxj, Wang:2023zxj}, as well as some experimental values \cite{ParticleDataGroup:2024cfk,DASP:1978dns}.}
	\label{tab: parameter}
	\setlength{\tabcolsep}{3pt}
	\centering 
	\begin{tabular}{ccc}
		\hline \hline
		States    &Mass (MeV) &$\Gamma$ (MeV) \\ \hline
		$\psi(3770)$  &$3773.7\pm0.7$~\cite{ParticleDataGroup:2024cfk} &$27.2\pm1.0$~\cite{ParticleDataGroup:2024cfk} \\   
		$\psi(4040)$  &$4040\pm4$~\cite{ParticleDataGroup:2024cfk}  &$84\pm12$~\cite{ParticleDataGroup:2024cfk}  \\
		$\psi(4160)$  &$4159\pm22$~\cite{DASP:1978dns} &$78\pm22$~\cite{DASP:1978dns} \\
		$\psi(4220)$  &4222 &44 \\
		$\psi(4380)$  &4389 &80 \\
		$\psi(4415)$  &4414 &33 \\
		$\psi(4500)$  &4509 &50 \\
		\hline \hline 	
	\end{tabular}
\end{table}

	As discussed in Ref.~\cite{LHCb:2013ywr}, the sideband background contributes a smooth curve to the mass spectrum. To simulate this background contribution, we use a polynomial in \( m_{\mu^+\mu^-} \), which is expressed as,
	
	\begin{align}	
		\mathcal{P}_{\mathrm{back}}(m_{\mu^+\mu^-})=a+bm_{\mu^+\mu^-}+cm_{\mu^+\mu^-}^2.
	\end{align}	
	By fitting the background given in Fig. 1 of Ref.~\cite{LHCb:2013ywr}, we obtain $a=-36.42$, $b=0.03\text{ MeV}^{-1}$, and $c=4.14\times10^{-6}\text{ MeV}^{-2}$.

	To directly compare with the experimental data, the detection and reconstruction efficiency must be included. This efficiency should vary smoothly across the entire dimuon invariant mass range, with a drop near the edges of the phase space. Specifically, when \( m_{\mu^+\mu^-} \) is close to \( \Delta = m_B - m_K \), the kaon momentum becomes very small, leading to a lower detection efficiency for low-momentum kaons. Therefore, we use an efficiency curve parameterized by
	
	\begin{align}	
		\varepsilon(m_{\mu^+\mu^-})\propto1-\frac{1}{1+\kappa(\Delta-m_{\mu^+\mu^-})/\Delta},	
	\end{align}		
	which accounts for the variation of detection efficiency across the mass spectrum.
	
	As indicated in Ref.~\cite{LHCb:2013ywr}, the efficiency decreases by approximately $20\%$ in the region between the $J/\psi$ mass and $4600$ MeV. From this, we can estimate the parameter \( \kappa \simeq 88 \). After incorporating the efficiency correction, the observed event distribution is given by
	\begin{align}	
		\frac{dN}{dm_{\mu^+\mu^-}}\propto\mathcal{P}_{\mathrm{back}}(m_{\mu^+\mu^-})+\mathcal{P}_{\mathrm{sig}}\times\varepsilon(m_{\mu^+\mu^-}).
	\end{align}

With the above preparation and considering the contributions of higher charmonium states in the range of $3.7 - 4.5$~GeV from the theoretical predictions~\cite{Wang:2019mhs} listed in Table~\ref{tab: parameter}, we perform an analysis of the di-muon invariant mass spectrum from the $B^+ \to K^+ \mu^+ \mu^-$ process. This analysis aims to refine the resonance parameters of the $\psi(4160)$ and verify the unquenched charmonium spectrum.

We fit seven resonances in the di-muon invariant mass spectrum, in which the resonance parameters of all charmonium states are used, incorporating both experimental data and theoretical predictions. The resulting fitted parameters are listed in Table~\ref{tab: fitting parameters}.

\begin{figure}[htbp]\centering
	\includegraphics[width=0.45\textwidth]{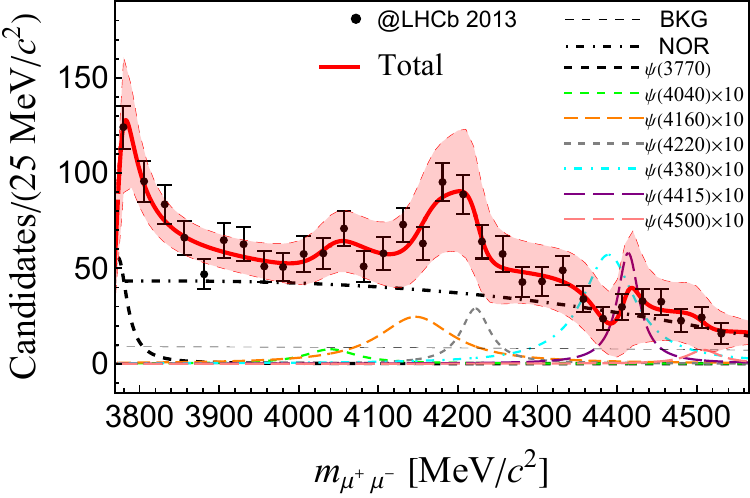}
	\caption{The black dots with error bands represent experimental data from the LHCb Collaboration \cite{LHCb:2013ywr}. The red curve, accompanied by a shaded band, represents the total contribution, including uncertainties. The background and non-resonance components are depicted by thin dashed and dot-dashed lines, respectively. The dashed lines show the individual contributions from the $\psi(3770)$, $\psi(4040)$, $\psi(4160)$, $\psi(4220)$, $\psi(4380)$, $\psi(4415)$, and $\psi(4500)$. To enhance the visibility of the contributions from the last six states, their amplitudes are scaled by a factor of ten.}\label{fig1}
\end{figure}

In Fig.~\ref{fig1}, the line shape is well described across the entire energy range from $4$ to $4.5$ GeV within the experimental error range, and the peaks around 4190 MeV can be naturally explained by the interference effects between the $\psi(4160)$ and $\psi(4220)$. This indicates that our predictions for the highly excited charmonium states—$\psi(4040)$, $\psi(4160)$, $\psi(4220)$, $\psi(4380)$, $\psi(4415)$, and $\psi(4500)$—under the unquenched framework are reasonable. This finding should encourage experimental studies to pay closer attention to the unquenched charmonium spectrum, as well as to the interference effect between the $\psi(4160)$ and $\psi(4220)$, in future analyses.

\section{Reinforcing this scenario by $e^+e^- \to D_s \bar{D}_s^*$}
Due to the interference effects between the $\psi(4160)$ and $\psi(4220)$, the structure around 4.2 GeV in many processes, can reveal a charmonium signal peak with a mass ranging from 4140 MeV to 4200 MeV. If we only consider the contribution of the $\psi(4160)$ and ignore the presence of the $\psi(4220)$ around 4.2 GeV, it could lead to a misleading understanding of the resonance parameters of the $\psi(4160)$.

Since it is difficult to isolate the contribution of the lower mass state of the $\psi(4160)$ in most processes, we choose the $e^+e^- \to D_s \bar{D}_s^*$ ~\cite{CLEO:2008ojp, Belle:2010fwv} process to study this issue, as it may be the only process capable of directly confirming the lower mass of the $\psi(4160)$ at present. We revisit the resonance parameters of the $\psi(4160)$ by considering the interference effects between the $\psi(4160)$ and $\psi(4220)$ within the framework of the unquenched charmonium spectrum. In the $e^+e^- \to D_s \bar{D}_s^*$ reaction~\cite{CLEO:2008ojp, Belle:2010fwv}, the cross section is expressed as:

\begin{align}
	\sigma(s)=|\text{BW}_1(s)+\text{BW}_2(s)e^{i\phi}|^2,
\end{align}
where the relativistic Breit-Wigner (BW) amplitude for the resonance \( \psi \to D_s \bar{D}_s^* \) is given by:
\begin{align}
	\mathrm{BW}(s)=\frac{\sqrt{12\pi\Gamma^{ee}_\psi\Gamma_{\mathrm{tot}}\mathcal{B}(\psi \to D_s \bar{D}_s^*)}}{s-M^2+iM\Gamma_{\mathrm{R}}}\sqrt{\frac{\mathrm{PS}(\sqrt{s})}{\mathrm{PS}(M)}},
\end{align}
where $M$, $\Gamma_{\mathrm{R}}$, $\Gamma^{ee}_\psi$, and $\mathcal{B}(\psi \to D_s \bar{D}_s^*)$ are the mass, total width, electronic partial width, and branching fraction of the corresponding resonance, respectively. Additionally, $\sqrt{\text{PS}(\sqrt{s})/\text{PS}(M)}$ is the two-body phase space factor.

\begin{table}[tbp]
	\renewcommand\arraystretch{1.5}
	\caption{The parameter values obtained from fitting the experimental data are as follows. The factors $f_i$ (in units of MeV) are chosen to ensure that the resonance amplitudes have the same dimensions as the non-resonance contribution. The phases $\delta_i$ (in radians) correspond to the seven $\psi$ states: $\psi(3770)$, $\psi(4040)$, $\psi(4160)$, $\psi(4220)$, $\psi(4380)$, $\psi(4415)$, and $\psi(4500)$, listed in succession.} 
	\label{tab: fitting parameters}
	\centering 
	\begin{tabular}{c|c|c|c}
		\hline \hline
		Parameters &  Value &	Parameters &  Value \\ 
		\hline
		$f_1~\text{(MeV)}$  ~~&~~$46.37\pm2.52$  ~~&~~$\delta_1~\text{(rad)}$ ~~&~~$0.95\pm0.05$\\			
		$f_2~\text{(MeV)}$  ~~&~~$4.83\pm0.33$   ~~&~~$\delta_2~\text{(rad)}$  ~~&~~$2.30\pm0.17$\\					
		$f_3~\text{(MeV)}$  ~~&~~$7.12\pm0.56$   ~~&~~$\delta_3~\text{(rad)}$  ~~&~~$1.67\pm0.11$\\		
		$f_4~\text{(MeV)}$  ~~&~~$8.85\pm0.56$   ~~&~~$\delta_4~\text{(rad)}$  ~~&~~$4.36\pm0.32$\\			
		$f_5~\text{(MeV)}$  ~~&~~$9.87\pm0.74$   ~~&~~$\delta_5~\text{(rad)}$  ~~&~~$5.66\pm0.42$\\	
		$f_6~\text{(MeV)}$  ~~&~~$9.29\pm0.49$   ~~&~~$\delta_6~\text{(rad)}$  ~~&~~$2.74\pm0.20$\\			
		$f_7~\text{(MeV)}$  ~~&~~$3.57\pm0.26$   ~~&~~$\delta_7~\text{(rad)}$  ~~&~~$5.00\pm0.30$\\	
		\hline
		\multicolumn{4}{c}{$\chi^2/\text{d.o.f.}=0.90$}   \\ \hline \hline 	
	\end{tabular}
\end{table}

\begin{figure}[tbp]\centering
	\includegraphics[width=0.45\textwidth]{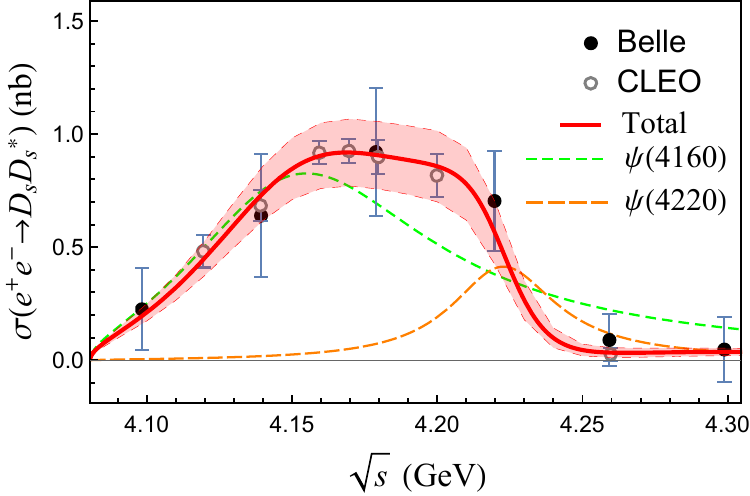}
	\caption{The black and hollow dots with error bands represent the experimental data from the Belle and CLEO Collaborations \cite{CLEO:2008ojp, Belle:2010fwv}, respectively. The dashed lines show the individual contributions from the $\psi(4160)$ and $\psi(4220)$ resonances, respectively. The red curve with a band represents the total contribution and uncertainties.}
	\label{fig2}
\end{figure}

In the analysis, we use free resonance parameters for the $\psi(4160)$, while the resonance parameters for the $\psi(4220)$ are listed in Table~\ref{tab: parameter}. From the analysis shown in Fig.~\ref{fig2}, a lower mass of the $\psi(4160)$ around $4.140-4.160$ GeV is evident. This is consistent with the results from previous experiments~\cite{DASP:1978dns, Seth:2004py}, as shown in Fig.~\ref{fig0}. The resonance parameters of the $\psi(4160)$ from the best fit are 
\begin{align}
	m_{\psi(4160)} =& 4145.76\pm5.48~\mathrm{MeV}, \nonumber\\\nonumber
	\Gamma_{\psi(4160)} =& 104.83\pm23.71~\mathrm{MeV}.
\end{align} 
We also test four fitting scenarios by fixing the mass of the $\psi(4160)$ to 4140, 4150, 4160, and 4170 MeV and the corresponding results are summarized in Table~\ref{tab: DsDs}. One can see that when the mass of the $\psi(4160)$ exceeds 4170 MeV, the line shape no longer describes the data well, as evidenced by the large $\chi^2/\text{d.o.f.} = 1.88$ shown in Table~\ref{tab: DsDs}.

\begin{widetext}

\begin{table*}[hb]
	\renewcommand\arraystretch{1.5}
	\caption{The fitting parameters $m_{\psi(4160)}$, $\Gamma_{\psi(4160)}$, $\Gamma^{ee}_{\psi}$ and $\mathcal{B}_{\psi}$ represent mass, total width, di-lepton width and branch ratio of $\psi \to D_s \bar{D}_s^*$, respectively, and $\phi~\text{(rad)}$ is the phase between the resonance amplitudes associated with the  $\psi(4160)$ and $\psi(4220)$ in the $e^+e^- \to D_s \bar{D}_s^*$ cross section.}
	\label{tab: DsDs}
	\centering 
	\begin{tabular}{c|c|c|c|c|c}
		\hline \hline
		Parameters &Best fit &I  &II &III  &IV \\ 
		\hline 	
		$m_{\psi(4160)}$~(MeV)                                   ~~&~~ $4145.76\pm5.48$   ~~&~~4140 (fixed)       ~~&~~4150 (fixed)       ~~&~~4160 (fixed)       ~~&~~4170 (fixed) \\
		$\Gamma_{\psi(4160)}$~(MeV)                              ~~&~~$104.83\pm23.71$    ~~&~~$113.98\pm24.01$  ~~&~~$108.78\pm23.65$   ~~&~~$127.17\pm17.65$   ~~&~~$143.37\pm25.24$  \\
		
		$\Gamma^{ee}_{\psi(4160)}\mathcal{B}_{\psi(4160)}$~(eV)  ~~&~~$98.02\pm26.88$     ~~&~~$108.43\pm29.8$    ~~&~~$109.14\pm30.04$   ~~&~~$168.42\pm35.09$   ~~&~~$207.29\pm25.91$  \\ 
		$\Gamma^{ee}_{\psi(4220)}\mathcal{B}_{\psi(4220)}$~(eV)  ~~&~~$22.09\pm8.82$      ~~&~~$23.23\pm10.00$    ~~&~~$21.32\pm10.96$    ~~&~~$51.12\pm34.72$    ~~&~~$66.81\pm22.22$\\ 				
		$\phi~\text{(rad)}$                                      ~~&~~$2.91\pm0.27$       ~~&~~$2.90\pm0.23$      ~~&~~$3.01\pm0.30$      ~~&~~$3.35\pm0.23$      ~~&~~$3.25\pm0.14$ \\ 			
		\hline 
		$\chi^2/\text{d.o.f.}$                                   ~~&~~0.22                ~~&~~$0.31$             ~~&~~$0.26$             ~~&~~$0.79$             ~~&~~$1.88$\\			
		\hline \hline 	
	\end{tabular}
\end{table*}	
    
\end{widetext}

These results suggest that the resonance parameters of the $\psi(4160)$ require careful reconsideration, with a lower mass for the $\psi(4160)$ appearing to be a more plausible scenario. This finding should motivate experimental investigations to focus more closely on the unquenched charmonium spectrum, particularly the interference effects between the $\psi(4160)$ and $\psi(4220)$, in future analyses. Precisely determining the mass of the $\psi(4160)$ is essential for advancing charmonium spectroscopy.

\section{Summary}
A puzzling increase in the measured mass of the $\psi(4160)$ presents a significant challenge to current theoretical models of hadron spectroscopy. This work indicates that this issue arises from analyses limited to the quenched charmonium mass spectrum, which is outdated in the era of high-precision hadron spectroscopy. In the past two decades, with the discovery of numerous new hadronic states, the unquenched effect has become increasingly important. Under the unquenched picture \cite{Wang:2019mhs}, there are six vector charmonium states in the $4 \sim 4.5$ GeV range: $\psi(4040)$, $\psi(4160)$, $\psi(4220)$, $\psi(4380)$, $\psi(4415)$, and $\psi(4500)$, a significant departure from the three states predicted in the quenched picture \cite{Eichten:1979ms}.

Using the unquenched charmonium assignment, we reevaluate the resonance parameters of the $\psi(4160)$ by combining the di-muon invariant mass spectrum of $B^+ \to K^+ \mu^+ \mu^-$ with nearby contributions from higher charmonium $\psi(4220)$.  We find that previous experimental analyses have overestimated the mass of the  $\psi(4160)$ \cite{BES:2007zwq,LHCb:2013ywr,BESIII:2023wsc}, a conclusion further supported by examining $e^+e^- \to D_s \bar{D}_s^*$, which yields a mass for the $\psi(4160)$ of $4145.76 \pm 4.48$ MeV.

This work not only addresses a longstanding puzzle regarding the measured mass shift of the $\psi(4160)$ but also provides important insights for hadron spectroscopy. Additionally, our findings have implications for the study of new physics. The decay $B^+ \to K^+ \mu^+ \mu^-$ is closely related to the measurement of $R_K$ \cite{LHCb:2021trn}, which is crucial for investigating physics beyond the Standard Model. An accurate determination of the contributions from intermediate charmonium states is essential for improving the precision of $R_K$ and for interpreting potential new physics signals.

\section*{ACKNOWLEDGMENTS}

This work is supported by the National Natural Science Foundation of China under Grant Nos. 12335001, 12247101, 12447124, and 12405088, the ‘111 Center’ under Grant No. B20063, the Natural Science Foundation of Gansu Province (No. 22JR5RA389), the fundamental Research Funds for the Central Universities, and the project for top-notch innovative talents of Gansu province. J.Z.W. is also supported by the National Postdoctoral Program for Innovative Talent.
T. P. is supported by the Gansu Province Postgraduate Innovation Star Program No. 2025CXZX-043.


\begin{thebibliography}{99}
	\bibitem{E598:1974sol}
	J.~J.~Aubert \textit{et al.} [E598],
	Experimental Observation of a Heavy Particle $J$,
	\href{https://journals.aps.org/prl/abstract/10.1103/PhysRevLett.33.1406}{Phys. Rev. Lett. \textbf{33}, 1404-1406 (1974)}.
	
	\bibitem{SLAC-SP-017:1974ind}
	J.~E.~Augustin \textit{et al.} [SLAC-SP-017],
	Discovery of a Narrow Resonance in $e^+ e^-$ Annihilation,
	\href{https://journals.aps.org/prl/abstract/10.1103/PhysRevLett.33.1406}{Phys. Rev. Lett. \textbf{33}, 1406-1408 (1974)}.
	doi:10.1103/PhysRevLett.33.1406

    \bibitem{ParticleDataGroup:2024cfk}
    S.~Navas \textit{et al.} [Particle Data Group],
    Review of particle physics,
    \href{https://doi.org/10.1103/PhysRevD.110.030001}{Phys. Rev. D \textbf{110}, 030001 (2024)}.
	
	\bibitem{DASP:1978dns}
	R.~Brandelik \textit{et al.} [DASP],
	Total Cross-section for Hadron Production by $e^+ e^-$ Annihilation at Center-of-mass Energies Between 3.6-{GeV} and 5.2-{GeV},
	\href{https://linkinghub.elsevier.com/retrieve/pii/0370269378908079}{
		Phys. Lett. B \textbf{76}, 361 (1978)}.
	
	
	\bibitem{Seth:2004py}
	K.~K.~Seth,
	Alternative analysis of the R measurements: Resonance parameters of the higher vector states of charmonium,
	\href{https://journals.aps.org/prd/abstract/10.1103/PhysRevD.72.017501}{Phys. Rev. D \textbf{72}, 017501 (2005)}.
	
	
	\bibitem{BES:2007zwq}
	M.~Ablikim \textit{et al.} [BES],
	Determination of the $\psi(3770)$, $\psi(4040)$, $\psi(4160)$ and $\psi(4415)$ resonance parameters,
	eConf \textbf{C070805}, 02 (2007)
	\href{https://www.sciencedirect.com/science/article/pii/S0370269308000397?via%3Dihub}{doi:10.1016/j.physletb.2007.11.100}.
	
	
	\bibitem{LHCb:2013ywr}
	R.~Aaij \textit{et al.} [LHCb],
	Observation of a resonance in $B^+ \to K^+ \mu^+\mu^-$ decays at low recoil,
	\href{https://journals.aps.org/prl/abstract/10.1103/PhysRevLett.111.112003}{Phys. Rev. Lett.  \textbf{111}, no.11, 112003 (2013)}.
	
	\bibitem{BESIII:2023wsc}
	M.~Ablikim \textit{et al.} [BESIII],
	Precise Measurement of the $e^+e^- \to D_s^{*+}D_s^{*-}$ Cross Sections at Center-of-Mass Energies from Threshold to 4.95~GeV,
	\href{https://journals.aps.org/prl/abstract/10.1103/PhysRevLett.131.151903}{Phys. Rev. Lett. \textbf{131}, no.15, 151903 (2023)}.
	
		\bibitem{He:2014xna}
		L.~P.~He, D.~Y.~Chen, X.~Liu, and T.~Matsuki,
		Prediction of a missing higher charmonium around 4.26 GeV in $J/\psi$ family,
		\href{https://link.springer.com/article/10.1140/epjc/s10052-014-3208-5}{Eur. Phys. J. C \textbf{74}, 3208 (2014)}.
		
		\bibitem{Chen:2014sra}
		D.~Y.~Chen, X.~Liu, and T.~Matsuki,
		Observation of $e^+e^-\to \chi_{c0}\omega$ and missing higher charmonium $\psi(4S)$,
		\href{https://journals.aps.org/prd/abstract/10.1103/PhysRevD.91.094023}{Phys. Rev. D \textbf{91}, 094023 (2015)}.
		
		\bibitem{Chen:2015bma}
		D.~Y.~Chen, X.~Liu, and T.~Matsuki,
		Search for missing $\psi(4S)$ in the $e^+e^-\to \pi^+\pi^-\psi(2S)$ process,
		\href{https://journals.aps.org/prd/abstract/10.1103/PhysRevD.93.034028}{Phys. Rev. D \textbf{93}, 034028 (2016)}.
		
		\bibitem{Chen:2017uof}
		D.~Y.~Chen, X.~Liu, and T.~Matsuki,
		Interference effect as resonance killer of newly observed charmoniumlike states $Y(4320)$ and $Y(4390)$,
		\href{https://link.springer.com/article/10.1140/epjc/s10052-018-5635-1}{Eur. Phys. J. C \textbf{78}, 136 (2018)}.
	
    \bibitem{Wang:2019mhs}
    J.~Z.~Wang, D.~Y.~Chen, X.~Liu, and T.~Matsuki,
    Constructing $J/\psi$ family with updated data of charmoniumlike $Y$ states,
    \href{https://doi.org/10.1103/PhysRevD.99.114003}{Phys. Rev. D \textbf{99}, no.11, 114003 (2019)}.
	
	
	\bibitem{Wang:2022jxj}
	J.~Z.~Wang and X.~Liu,
	Confirming the existence of a new higher charmonium \ensuremath{\psi}(4500) by the newly released data of $e^+e^-\to K^+K^-J/\psi$,
	\href{https://journals.aps.org/prd/abstract/10.1103/PhysRevD.107.054016}{Phys. Rev. D \textbf{107}, 054016 (2023)}.
	
	\bibitem{Wang:2023zxj}
	J.~Z.~Wang and X.~Liu,
	Identifying a characterized energy level structure of higher charmonium well matched to the peak structures in $e^+e^-\to \pi^+D^0D^{*-}$,
	\href{https://www.sciencedirect.com/science/article/pii/S0370269324000157}{Phys. Lett. B \textbf{849}, 138456 (2024)}.
	
	
	
	\bibitem{Peng:2024xui}
	T.~C.~Peng, Z.~Y.~Bai, J.~Z.~Wang and X.~Liu,
	How higher charmonia shape the puzzling data of the $e^+e^-\to \eta J/\psi$ cross section,
	\href{https://journals.aps.org/prd/abstract/10.1103/PhysRevD.109.094048}{Phys. Rev. D \textbf{109}, no.9, 094048 (2024)}.
	
	
	\bibitem{BaBar:2005hhc}
	B.~Aubert \textit{et al.} [BaBar],
	Observation of a broad structure in the $\pi^+ \pi^- J/\psi$ mass spectrum around 4.26 GeV/$c^2$,
	\href{https://journals.aps.org/prl/abstract/10.1103/PhysRevLett.95.142001}{Phys. Rev. Lett. \textbf{95}, 142001 (2005)}.
	
	\bibitem{BESIII:2016bnd}
	M.~Ablikim \textit{et al.} [BESIII],
	Precise measurement of the $e^+e^-\to \pi^+\pi^-J/\psi$ cross section at center-of-mass energies from 3.77 to 4.60 GeV,
	\href{https://journals.aps.org/prl/abstract/10.1103/PhysRevLett.118.092001}{Phys. Rev. Lett. \textbf{118}, 092001 (2017)}.
	
	\bibitem{BESIII:2020oph}
	M.~Ablikim \textit{et al.} [BESIII],
	Study of the process $e^+e^-\to\pi^0\pi^0 J/\psi$ and neutral charmonium-like state $Z_c(3900)^0$,
	\href{https://journals.aps.org/prd/abstract/10.1103/PhysRevD.102.012009}{Phys. Rev. D \textbf{102}, 012009 (2020)}.
	
	
	\bibitem{BaBar:2006ait}
	B.~Aubert \textit{et al.} [BaBar]
	Evidence of a broad structure at an invariant mass of 4.32 $\text{GeV}/c^{2}$ in the reaction $e^{+} e^{-} \to \pi^{+} \pi^{-} \psi(2S)$ measured at BaBar,
	\href{https://journals.aps.org/prl/abstract/10.1103/PhysRevLett.98.212001}{Phys. Rev. Lett. \textbf{98}, 212001 (2007)}.
	
	
	\bibitem{BESIII:2017tqk}
	M.~Ablikim \textit{et al.} [BESIII],
	Measurement of $e^{+}e^{-}\rightarrow \pi^{+}\pi^{-}\psi(3686)$ from 4.008 to 4.600 GeV and observation of a charged structure in the $\pi^{\pm}\psi(3686)$ mass spectrum,
	\href{https://journals.aps.org/prd/abstract/10.1103/PhysRevD.96.032004}{Phys. Rev. D \textbf{96}, 032004 (2017)}.
	
	\bibitem{BESIII:2016adj}
	M.~Ablikim \textit{et al.} [BESIII],
	Evidence of Two Resonant Structures in $e^+ e^- \to \pi^+ \pi^- h_c$,
	\href{https://journals.aps.org/prl/abstract/10.1103/PhysRevLett.118.092002}{Phys. Rev. Lett. \textbf{118}, 092002 (2017)}.
	
	\bibitem{BESIII:2019gjc}
	M.~Ablikim \textit{et al.} [BESIII],
	Cross section measurements of $e^+ e^-\to\omega\chi_{c0}$ form $\sqrt{s}=$ 4.178 to 4.278 GeV,
	\href{https://journals.aps.org/prd/abstract/10.1103/PhysRevD.99.091103}{Phys. Rev. D \textbf{99}, 091103 (2019)}.
	
	\bibitem{BESIII:2022joj}
	M.~Ablikim \textit{et al.} [(BESIII), and BESIII],
	Observation of the Y(4230) and a new structure in $e^+e^-\to K^+K^-J/\psi$,
	\href{https://iopscience.iop.org/article/10.1088/1674-1137/ac945c}{Chin. Phys. C \textbf{46}, 111002 (2022)}.
	
	
	\bibitem{BESIII:2018iea}
	M.~Ablikim \textit{et al.} [BESIII],
	Evidence of a resonant structure in the $e^+e^-\to \pi^+D^0D^{*-}$ cross section between 4.05 and 4.60 GeV,
	\href{https://journals.aps.org/prl/abstract/10.1103/PhysRevLett.122.102002}{Phys. Rev. Lett. \textbf{122}, 102002 (2019)}.
	
	
	
	\bibitem{BESIII:2023tll}
	M.~Ablikim \textit{et al.} [BESIII],
	Measurement of $e^+e^-\to \eta J/\psi$ cross section from s=3.808 GeV to 4.951 GeV,
	\href{https://journals.aps.org/prd/abstract/10.1103/PhysRevD.109.092012}{Phys. Rev. D \textbf{109}, no.9, 092012 (2024)}.
	
	
	\bibitem{ARGUS:1990hfq}
	H.~Albrecht \textit{et al.} [ARGUS],
	Search for Hadronic $b \to u$ Decays,
	\href{https://www.sciencedirect.com/science/article/abs/pii/037026939091293K}{Phys. Lett. B \textbf{241}, 278-282 (1990)}.
	
	\bibitem{CLEO:2008ojp}
	D.~Cronin-Hennessy \textit{et al.} [CLEO],
	Measurement of Charm Production Cross Sections in $e^+e^-$ Annihilation at Energies between 3.97 and 4.26-GeV,
	\href{https://journals.aps.org/prd/abstract/10.1103/PhysRevD.80.072001}{Phys. Rev. D \textbf{80}, 072001 (2009)}.
	
	\bibitem{Belle:2010fwv}
	G.~Pakhlova \textit{et al.} [Belle],
	Measurement of $e^+e^-\to D_s^{(*)+} D_s^{(*)-}$ cross sections near threshold using initial-state radiation,
	\href{https://journals.aps.org/prd/abstract/10.1103/PhysRevD.83.011101}{Phys. Rev. D \textbf{83}, 011101 (2011)}.
	
	
    \bibitem{Eichten:1979ms}
    E.~Eichten, K.~Gottfried, T.~Kinoshita, K.~D.~Lane and T.~M.~Yan,
    Charmonium: Comparison with Experiment,
    \href{https://journals.aps.org/prd/abstract/10.1103/PhysRevD.21.203}{Phys. Rev. D \textbf{21}, 203 (1980)}.

    \bibitem{LHCb:2021trn}
    R.~Aaij \textit{et al.} [LHCb],
    Test of lepton universality in beauty-quark decays,
    \href{https://www.nature.com/articles/s41567-021-01478-8}{Nature Phys. \textbf{18}, no.3, 277-282 (2022)}.

\end{thebibliography}
\end{document}